\documentclass[prb,twocolumn,showpacs,preprintnumbers,amsmath]{revtex4}
\usepackage{dcolumn}
\usepackage{bm}
\usepackage{graphicx}
\usepackage{color}
\usepackage{bm}
\begin{document}

\title{The dynamics of magnetization in phase separated manganite around half doping: a case study for Pr$_{0.5}$Sr$_{0.5}$Mn$_{0.925}$Ga$_{0.075}$O$_{3}$}

\author{A. K. Pramanik}\altaffiliation{Present address: Leibniz Institute for Solid State and Materials Research (IFW) Dresden, D-01171 Dresden, Germany.}\affiliation{UGC-DAE Consortium for Scientific Research, University Campus, Khandwa Road, Indore-452001, M. P., India.}
\author{A. Banerjee}\affiliation{UGC-DAE Consortium for Scientific Research, University Campus, Khandwa Road, Indore-452001, M. P., India.}
 
\begin{abstract}
We investigate the dynamics of magnetization in the phase separated (PS) state after introducing the quenched disorder at the Mn-site of a manganite around half doping. The compound, Pr$_{0.5}$Sr$_{0.5}$Mn$_{0.925}$Ga$_{0.075}$O$_{3}$, exhibits PS with the coexistence of ferromagnetic (FM) and antiferromagnetic (AFM) clusters where the size of the FM clusters is substantially reduced due to the disorder introduced by nonmagnetic Ga substitution. At low temperature, the system develops a new magnetic anomaly, which is marked by a peak in the zero field cooled magnetization. Detailed study of linear as well as nonlinear ac susceptibilities coupled with dc magnetization indicates that this peak arises due to the thermal blocking of nanometer size FM clusters demonstrating superparamagnetic behavior. The system, however, exhibits slow magnetic relaxation, aging effect, memory effect in both field cooled and zero field cooled magnetization below the blocking temperature.  These imply the presence of collective behavior induced by the interaction between the clusters. Moreover, the magnetic relaxation measured with positive and negative temperature excursions exhibits asymmetric response suggesting that the dynamics in this phase separated system is accounted by the hierarchical model rather than the droplet model which are commonly used to describe the similar collective dynamics in glassy system.      
\end{abstract}
\maketitle
\section {Introduction}
The phase separation (PS) and the effect of quenched disorder have been studied intensively in the hole doped manganites with chemical formula $R_{1-x}$$A_{x}$MnO$_3$, where $R$ and $A$ represent the trivalent rare earth and divalent alkaline earth element, respectively.\cite{dago} At low temperature ($T$), the dynamics of the PS systems are quite interesting as they exhibit various fascinating experimental features similar to spin glass (SG) system \cite{mydosh,binder} like, long time relaxation and aging effect,\cite{lopez,freitas} rejuvenation and persistent memory effect,\cite{levy} memory and aging effect,\cite{rivadulla} etc. It is commonly believed that the glassy behavior in the PS systems originates due to the strong interactions among the magnetic clusters which induces the collectivity and slow dynamics in the system, yet, whether their dynamics is consistent with the classical SG systems or not is under debate.\cite{dago,rivadulla} Sometimes, these self-generated clusters are also thermally blocked at low temperature yielding superparamagnetic (SPM) behavior.\cite{sunilprl}
It remains a challenging task to discern between the SG and SPM behavior due to their many shared features, however, their dynamics are quite different in terms of relaxation time which is \textit{critically} slow in former case compared to slow but non-critical in later systems. Hence, a serious endeavor in experimental investigation is required to characterize the low temperature magnetic state in the PS compounds. In this context, the non-linear susceptibility turns out to be rather useful tool because it leaves distinct signatures for SG vis-$\grave{a}$-vis SPM like systems.\cite{sunilprl,ashna,akm,ranganathan,bitoh,suzuki,psmo-nano}       

Here we investigate the low temperature dynamics of PS manganite using a representative ceramic compound Pr$_{0.5}$Sr$_{0.5}$Mn$_{0.925}$Ga$_{0.075}$O$_{3}$ (PSMGO) by means of dc magnetization ($M$), linear and nonlinear ac susceptibility ($\chi$), magnetic relaxation and magnetic memory measurements. The parent compound Pr$_{0.5}$Sr$_{0.5}$MnO$_{3}$ (PSMO) is a half doped medium bandwidth manganite. On cooling from the room temperature, it shows second-order paramagnetic (PM) to ferromagnetic (FM) transition followed by first-order FM to antiferromagntic (AFM) transition where the later is shown to be kinetically arrested.\cite{tomioka,psmo-crit,psmo-Tn} The PSMO exhibits PS effect where the coexistence of FM and AFM clusters arise just below the PM-FM transition temperature $T_C$ and persists down to low temperature.\cite{psmo-ps} The introduction of quenched disorder in PSMO in the form of nonmagnetic Ga substitution at Mn-site has very interesting consequences which suppresses the FM state and reduces the size of the FM clusters.\cite{psmo-ps} These results agree with the theoretical prediction by Imry and Ma who argued in favor of destabilization of long range order system in presence of the quenched disorder,\cite{imry} which have also wide ramifications on the observed Griffiths phase like behavior in PSMO.\cite{psmo-gp} The compound under study i.e., PSMGO where 7.5\% Ga is substituted at the Mn-site of PSMO, is quite interesting because although its $\chi(T)$ resemble with SG system, the compound exhibits ferromagnetism, albeit highly inhomogeneous, and the FM cluster size is reduced significantly.\cite{psmo-ps} With this reduced size of FM clusters, the competition between thermal and magnetic energies of the clusters can give rise to various types of magnetic states (behaviors) at low temperature, which coupled with the presence of mixed interactions make this compound an ideal choice to study the dynamics and determine the low-$T$ magnetic state in the PS system.
 
Our results imply that the coexisting nanometer size FM clusters are thermally blocked at low temperature showing SPM behavior. Interestingly, this phenomenon is observed after a multiple magnetic transitions (PM-FM-AFM) which otherwise usually occurs after only FM or AFM transition. It is however noteworthy that below the blocking temperature $T_B$, the system exhibits long time magnetic relaxation, aging effect, zero field cooled (ZFC) and field cooled (FC) memory effect which are usually considered to be the benchmark of glassy dynamics. Though the first experimental evidence of low temperature blocked clusters in manganite has been shown for Pr$_{0.5}$Ca$_{0.5}$Mn$_{0.975}$Al$_{0.025}$O$_3$ (Ref. \onlinecite{sunilprl}) but the nature of clusters as well as the blocking mechanism appears to be different in present study. We attribute the observed experimental features in PSMGO to the intercluster interaction of the FM entities embedded in the AFM matrix.    
 
\maketitle\section{Experimental Details}
Polycrystalline sample Pr$_{0.5}$Sr$_{0.5}$Mn$_{0.925}$Ga$_{0.075}$O$_{3}$ has been prepared by the standard solid state ceramic route. This sample is a member of a series presented in Ref. \onlinecite{psmo-ps} where the sample preparation and characterization techniques have been given in detail. All the magnetic measurements have been performed in Quantum Design made vibrating sample magnetometer (VSM) and home made ac susceptometer.\cite{ashna-rsi} For the magnetic relaxation measurements, the sample has been cooled in zero field to the desired temperatures. After proper thermal stabilization a magnetic field ($H$) is applied and magnetization has been measured as a function of time ($t$). For the memory measurements, protocol has been described at the respective places in text.   

\begin{figure}
	\centering
		\includegraphics[width=7cm]{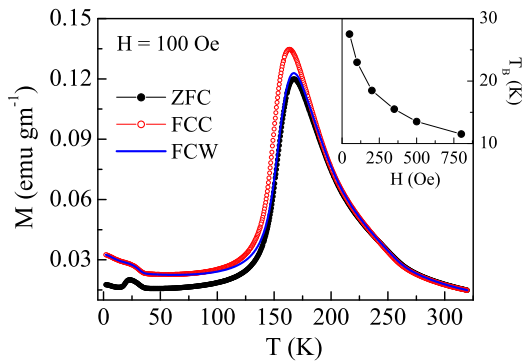}
	\caption{(Color online) The dc magnetization measured in 100 Oe following the ZFC, FCC and FCW protocols (defined in text) has been plotted against temperature for Pr$_{0.5}$Sr$_{0.5}$Mn$_{0.925}$Ga$_{0.075}$O$_{3}$. The inset shows the variation of peak temperature as a function of applied field.}
	\label{fig:Fig1}
\end{figure} 

\section{Results and discussions} 
Fig. 1 presents the dc magnetization measured in 100 Oe following the zero field cooled (ZFC), field cooled cooling (FCC) and field cooled warming (FCW) protocols. With lowering the temperature, magnetization measured in these three protocols initially increases and then decreases till about 50 K, showing a peak. The behavior of $M(T)$ appears to mimic that for the SG systems, however, it has been shown that, similar to parent compound PSMO, the composition PSMGO also exhibits FM state which is highly suppressed and inhomogeneous due to disorder (Ga substitution).\cite{psmo-ps} The inflection points in d$M$/d$T$ plot are used to calculate the PM to FM transition temperature ($T_C$) and FM to AF transition temperature ($T_N$), which are found to be around 174 K and 152 K, respectively. A thermal hysteresis (TH) between the FCC and FCW branches of magnetization is evident in Fig. 1 which starts below the $T_C$ and persists down to low temperature. The TH in magnetization is normally observed around the first order transition due to the supercooling and superheating of the high-$T$ and low-$T$ phases, respectively. However, the observed TH in present study indicates the coexistence of FM and AFM clusters in wide temperature range. Moreover, there remains a significant difference between the FC and ZFC branches of magnetization at low temperature which arises due to the arrested transformation kinetics of the high-$T$ FM phase into the low-$T$ AFM phases similar to the parent compound.\cite{psmo-Tn,manekar,kranti} 

\begin{figure}
	\centering
		\includegraphics[width=7cm]{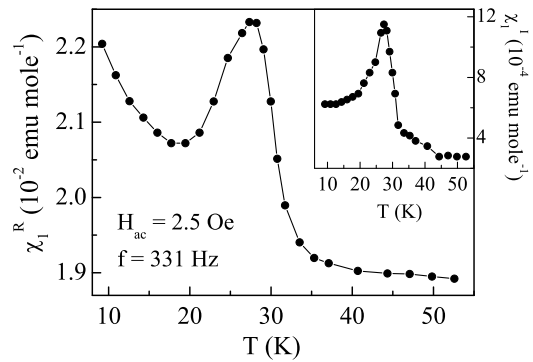}
	\caption{Real part of the first order ac-$\chi$ has been plotted as a function of temperature for Pr$_{0.5}$Sr$_{0.5}$Mn$_{0.925}$Ga$_{0.075}$O$_{3}$. In inset, the same has been shown for the imaginary part.}
	\label{fig:Fig2}
\end{figure}

An important observation is noticed at low temperature in Fig. 1 where the system exhibits a new magnetic state which is marked by a peak or cusp in $M_{ZFC}$ around 23 K and the enhanced splitting between the ZFC and FC magnetization below the peak. Note, that no TH is present between the FCC and FCW below the peak which discards the possibility of reentrant FM state at low temperature. Usually, this kind of experimental feature is associated with the SG like freezing or SPM like blocking mechanism. We have examined the effect of magnetic field on this peak. Inset of Fig. 1 shows that with increasing the measuring field peak in $M_{ZFC}$ shifts to lower temperature being consistent with SG or SPM behavior. However, the shifting of peak temperature with field here does not follow any functional form like de Almeida and Thouless (AT) line or Gabay and Toulouse (GT) line where the peak temperature varies $\propto$ $H^{2/3}$ or $\propto$ $H^{2}$, respectively. It can be mentioned that such lines were predicted for phase transition in the mean field theory for SG,\cite{mydosh,binder} and even have been observed for SPM system with interparticle interaction.\cite{psmo-nano}

To characterize the low temperature magnetic phase in this compound we have utilized the ac-$\chi$ which probes the dynamic nature of spin state in very low field. In general, magnetization can be written in terms of measuring field as:

\begin{eqnarray}
	M = M_0 + \chi_1 H + \chi_2 H^2 + \chi_3 H^3 + ......
\end{eqnarray}

where $M_0$ is the spontaneous magnetization, $\chi_1$ is the linear and $\chi_2$, $\chi_3$, ... are the nonlinear susceptibilities. Temperature dependence of the real part of first order ac-$\chi$ ($\chi_1^R$) measured in 2.5 Oe ac field and frequency ($f$) of 331 Hz is shown in main panel of Fig. 2. The $\chi_1^R$ exhibits a sharp peak in conformity with the low temperature anomaly observed ($\approx$ 23K) in dc-$M$ as shown in Fig. 1. The peak temperature ($T_B$) as calculated from the low field ac-$\chi$ is 27.5 K. The corresponding imaginary part ($\chi_1^I$), which signifies the magnetic loss, is plotted in the inset of Fig. 2 and shows peak at the same temperature of $\chi_1^R$. These results imply that peak in $M_{ZFC}$ at low temperature is associated with either SG or SPM like behavior. 

\begin{figure}
	\centering
		\includegraphics[width=8cm]{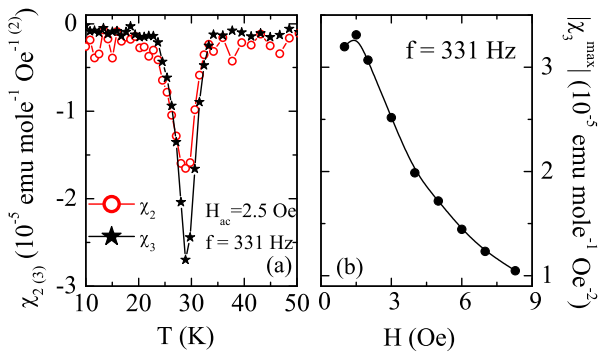}
	\caption{(Color online) (a) The real part of $\chi_2$ and $\chi_3$ measured in ac field 2.5 Oe and frequency 331 Hz are plotted against the temperature for Pr$_{0.5}$Sr$_{0.5}$Mn$_{0.925}$Ga$_{0.075}$O$_{3}$. (b) The maximum value of $\chi_3$ has been plotted as a function of applied ac field which clearly shows the noncritical behavior of $\chi_3$ with the applied field.}
	\label{fig:Fig3}
\end{figure}

Considering the fact that both SG and SPM have many common experimental features rendering the identification of these phases quite difficult, we have used nonlinear ac susceptibilities which are shown to be an effective probe to discern between various magnetic phases.\cite{sunilprl,psmo-ps,psmo-nano,ashna,akm,ranganathan,bitoh,sunilnsus,suzuki} In Fig. 3a, we have shown the temperature variation in second order ($\chi_2$) and third order ($\chi_3$) ac susceptibility measured in 2.5 Oe ac-field of frequency 331 Hz. Usually, $\chi_2$ becomes evident in experiment due to the symmetry breaking field which originates from either the spontaneous magnetization ($M_S$) or superimposed dc magnetic field.\cite{psmo-ps,akm,ranganathan} As for canonical SG, $M_S$ does not exist, therefore the $\chi_2$ is not evident in absence of dc field. However, there is a finite negative $\chi_2$ in Fig. 3a around $T_B$ which discards the possibility of SG like behavior in this compound. Similarly, $\chi_3 (T)$ also shows the characteristic negative peak around $T_B$ similar to SPM compound (Fig. 3a).\cite{ashna} To confirm the SPM behavior in PSMGO we have measured $\chi_3(T)$ at different applied fields because for SG, it has been conclusively shown that $\chi_3$ diverges as $H \rightarrow 0$.\cite{ashna,suzuki} In Fig. 3b we have plotted the maximum of $\chi_3$ as a function of the applied ac field. It is clear in figure that the $\chi_3$ shows noncritical behavior against the field, thus confirms the SPM behavior in this compound.         

\begin{figure}
	\centering
		\includegraphics[width=7cm]{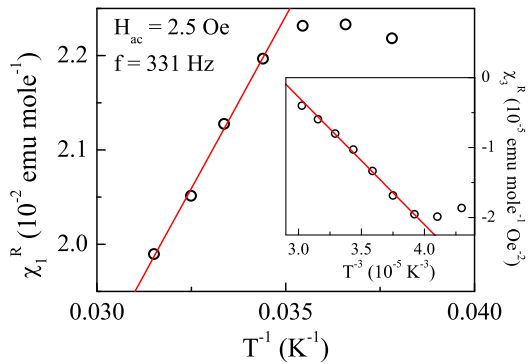}
	 \caption{(Color online) Temperature variation of $\chi_1^R$ above the blocking temperature. Straight line is a $T^{-1}$ fit to the $\chi_1$ data following Eq. 3. Inset: Similar temperature dependence of $\chi_3$ above the blocking temperature and the straight line is a $T^{-3}$ fit to $\chi_3$ data following Eq. 4.}	
	\label{fig:Fig4}
\end{figure}

We have further confirmed the SPM behavior of the magnetic clusters in PSMGO from the Wohlfarth's model which shows the magnetization of an assembly of particles can be given as:\cite{bitoh,wohl}

\begin{eqnarray}
	M = n\left\langle \mu \right\rangle L (\left\langle \mu \right\rangle H/k_BT)
\end{eqnarray}

where n is the number of particles per unit volume, $\left\langle \mu \right\rangle$ is the average moment of each magnetic particle, $k_B$ is the Boltzmann constant and L(x) is the Langevin function. After the expansion of L(x), $\chi_1$ and $\chi_3$ above the $T_B$ can be expressed as a function temperature as following:

\begin{eqnarray}
	\chi_1 = n \left\langle \mu \right\rangle^2/3k_B T = P_1/T
\end{eqnarray}

\begin{eqnarray}
	\chi_3 = (n\left\langle \mu \right\rangle/45)(\left\langle \mu \right\rangle/k_BT)^3 = P_3/T^3
\end{eqnarray}

The Eqs. 3 and 4 imply that for SPM, $\chi_1$ and $\chi_3$ will vary linearly above the $T_B$ as a function of $T^{-1}$ and $T^{-3}$, respectively. The main panel of Fig. 4 clearly shows that $\chi_1^R$ is linearly dependent on $T^{-1}$ above the $T_B$. Similarly, in inset of Fig. 3 the linear behavior of $\chi_3^R$ vs $T^{-3}$ is depicted above the $T_B$. Moreover, from the ratio of $P_3$ and $P_1$, we have estimated the value of $\left\langle \mu \right\rangle$ which comes out to be around 8.98 $\times$ $10^4$ $\mu_B$, where $\mu_B$ is the effective Bohr magneton. This large value of $\left\langle \mu \right\rangle$ in consistent with the SPM clusters as it consists of large number of spins compared to PM where the $\left\langle \mu \right\rangle$ is limited to only few $\mu_B$. Assuming that the clusters are spherical, their effective size is calculated from the $\left\langle \mu \right\rangle$ yielding the value around 15 nm. This small size of the FM clusters that are embedded within the AFM matrix in PSMGO is in agreement with the results in Ref. \onlinecite{psmo-ps}, and shows the significance of disorder created by nonmagnetic (Ga) substitution, which simply reduces the size of FM clusters without adding magnetic interactions in the system. The above results, however, constitute a conclusive proof in favor of SPM behavior of the FM clusters which are even though magnetically ordered at $T_C$, but due to their smaller size the moments fluctuate coherently, and they are thermally blocked at low temperature. It is more intriguing that for the compound Pr$_{0.5}$Sr$_{0.5}$Mn$_{0.925}$Ga$_{0.075}$O$_{3}$ where the mixed interactions in the form of FM and AFM are evident, nevertheless, the system exhibits SPM behavior.

\begin{figure}
	\centering
		\includegraphics[width=7cm]{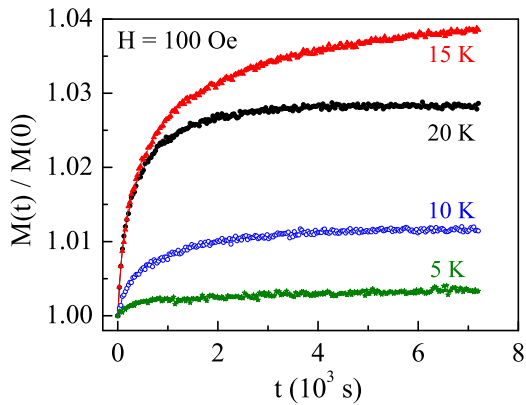}
	\caption{(Color online) Normalized magnetization measured at fixed temperature and field has been plotted as a function of time.}
	\label{fig:Fig5}
\end{figure}

After being confirmed the SPM behavior in PSMGO, we will now study the other phenomenon like magnetic relaxation, aging, memory effects, etc. which are typical to the PS systems. To look into the magnetic relaxation in this compound, we have measured time dependent magnetization, $M(t)$, in 100 Oe at 5, 10, 15 and 20 K for a time span of 7200 s, and the data are presented in Fig. 5 after normalized by $M(t=0)$. We find that magnetization increases continuously, however, it is evident in the figure that with the increasing temperature the relaxation rate ($S$) increases up to 15 K then decreases. The decreased $S$ at low temperatures is in agreement with the SPM system due to the reduced thermal fluctuation. At 20 K, $S$ again falls as this temperature is close to the $T_B$ in this measuring field (Fig. 1). Interestingly, the growth of $M(t)$ exhibits a nonlogarithmic time dependence at all temperatures unlike simple SPM system where the magnetization grows exponentially with time. This slow relaxation in this compound suggests the presence of correlated behavior which may arise due to the intercluster interaction and will be discussed in next sections. 

\begin{figure}
	\centering
		\includegraphics[width=6cm]{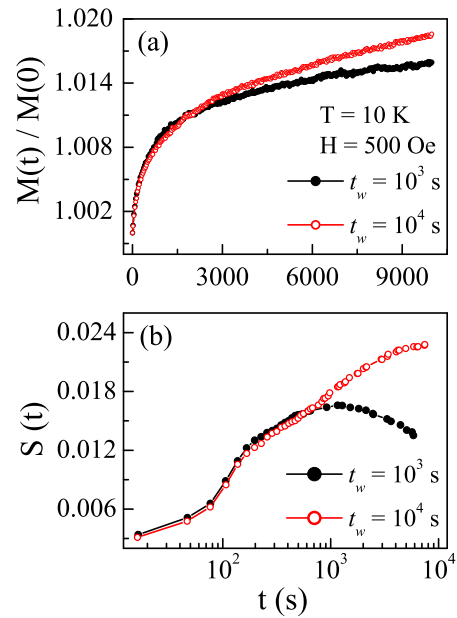}
	\caption{(Color Online) (a) The normalized magnetization vs time have been plotted at 10 K where the measuring field of 500 Oe is applied after $t_w$ = $10^3$ and $10^4$ s. (b) The magnetic relaxation rate $S(t)$ (defined in text) as extracted from the data in (a) have been plotted as a function of time for different wait time.}
	\label{fig:Fig6}
\end{figure}  

To probe the aging effect we have cooled the sample in zero field to 10 K ($< T_B$), and after waiting for certain amount of time ($t_w$) we have applied the field and recorded $M(t)$. In general, the aging occurs when the physical properties of a system evolve with time giving rise to a nonstationary phenomenon, and is known as a characteristic feature of correlated dynamics.\cite{granberg,akm1} The phenomenon of aging is, however, explained within the framework of existing models for SG. One is the droplet model suggested by Fisher and Huse\cite{fisher} and another is the hierarchical model as realized from the Parisi's solution of Sherrington-Kirkpatrick Hamiltonian.\cite{lefloch} The former model assumes the distribution of droplets of correlated spins which grow during $t_w$ to lower its energy by reducing the interface between the droplets of reverse character. The application of field after $t_w$ probes the correlated domains or droplets with time. On the other hand, the hierarchical model considers the metastable states, as separated by the finite barriers, are hierarchically organized in the free-energy landscape. This hierarchical structure is a function of temperature where the metastable states further split into the new (sub)states with decrease in temperature and merge into new state as temperature increases. In this approach, aging leads to distribution of population among the accessible states which evolves with time. However, the collective dynamics is required for the aging effect, hence many different systems like the ferrofluid with dipolar interactions\cite{jonsson} or the discontinuous multilayer with FM particles\cite{sahoo} are known to exhibit the aging effect.

The measured $M(t)$ with $t_w$ = $10^3$ and $10^4$ s for PSMGO have been presented in Fig. 6a after normalized by $M(t = 0)$. It is interesting to notice that the $M(t)$ varies with $t_w$ even though it is measured at same temperature and field which indicates that system ages during $t_w$. Note, that $M(t)$ does not exhibit any significant difference with $t_w$ up to $10^3$ s in present case, and the difference in $M(t)$ is observed with high $t_w$ only. Moreover, we have calculated $S(t)$[$= (1/H)(\partial M(t)/\partial \log(t))$] which are plotted in Fig. 6b for $t_w$ = $10^3$ and $10^4$ s. The $S(t)$ exhibits peak around $t = t_w$ in respective plot. Nonetheless, the aging behavior in this compound imply the presence of collective behavior in the system.

\begin{figure}
	\centering
		\includegraphics[width=7cm]{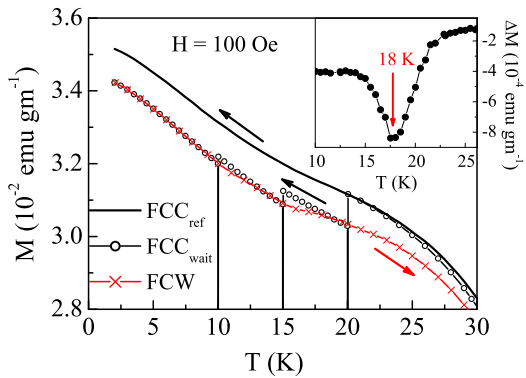}
	\caption{(Color online) Magnetic memory effect observed in Pr$_{0.5}$Sr$_{0.5}$Mn$_{0.925}$Ga$_{0.075}$O$_{3}$ sample. The main panel shows the memory effect during field cooled protocol (defined in text). The data has been collected in the measuring field of 100 Oe with the intermittent stop of 7200 s at 20, 15 and 10 K. The reference field cooled data has been shown as a line. Inset shows the memory effect as measured in zero field cooled protocol (see in text) where a stop of 7200 s has been given at 18 K.}
	\label{fig:Fig7}
\end{figure} 

The collective behavior often manifest itself through the magnetic memory effect. To test this, we have examined the FC memory effect which is shown to be characteristic feature of the interacting particle system.\cite{sun} For this purpose, we have cooled the sample in field (100 Oe) while collecting the magnetization data. During this FC process, we gave intermittent stop at temperature $T_i$ where the field has been withdrawn for a time $t_i$ at constant temperature thus allowing the system to relax, and then we resumed the FC process. We have selected $T_i$ = 20, 15 and 10 K and $t_i$ = 7200 s in our measurement. As evident in Fig. 7, the FC magnetization without `stop and wait' ($FC_{ref}$) rises continuously up to the lowest temperature. The FC magnetization with wait ($FC_{wait}$) shows lower value at each $T_i$ after the wait, however, with further cooling $FC_{wait}$ increases similar to $FC_{ref}$. This indicates that during $t_i$ the system ages through the distribution of particles among the available energy states. It remains interesting that upon subsequent reheating in same field, the magnetization ($FCW$) shows upturn at each $T_i$ and attempts to reaches the $FC_{wait}$ value at few kelvin above the $T_i$. In another way, the system remembers the waiting history during the FC process. This observed memory effect is quite noteworthy and implies the collective behavior in this compound arises from the intercluster interactions.\cite{sun}     
 
However, it has been subsequently shown that the noninteracting particles with the distribution of energy barriers can also give rise to the similar FC memory effect,\cite{sasakiprl,sasakiprb} therefore to confirm the intercluster interaction in this compound we have checked the ZFC memory effect. For this measurement, the sample has been cooled in zero field where during the cooling a temporary stop and wait is given at temperature $T_i$ ($<T_B$) for time $t_i$. During wait time the system relaxes toward equilibrium state. Then after reaching the low temperature, a field of 100 Oe has been applied and $M(T)$ ($ZFC_{wait}$) is subsequently recorded upon reheating. We have selected $T_i$ = 18 K and $t_i$ = 7200 s. For the system with interaction among the particles/clusters, the $ZFC_{wait}$ shows the lower value around $T_i$ than the ZFC magnetization ($ZFC_{ref}$) collected without stop and wait protocol. In other way, the $\Delta M = ZFC_{wait} - ZFC_{ref}$ exhibits dip around the $T_i$, signifying that the system has the memory of cooling history. The extracted $\Delta M (T)$ has been shown against the temperature in inset of Fig. 7. The $\Delta M$ clearly exhibits dip at $T_i$ which straightforwardly imply the presence of intercluster interaction in this compound. This observation of memory effect in the ZFC magnetization is quite noteworthy considering the fact that this experimental feature is commonly used as an identifying tool for the glassy behavior in many systems\cite{sasakiprb,chen,suzuki1} whereas the present system exhibits SPM behavior without an ambiguity. Nonetheless, this study shows that the collective behavior that causes slow dynamics is the prerequisite for different aging and memory effects which can even be evident in interactive SPM systems without having glass-like freezing. 
 
\begin{figure}[h]
	\centering
		\includegraphics[width=7cm]{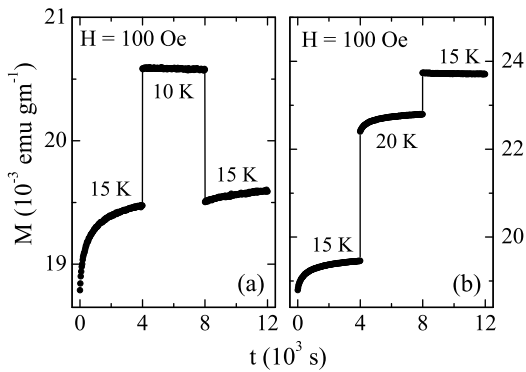}
	 \caption{(a) ZFC magnetic relaxation measured in 100 Oe at 15 K with an intermittent decrease in temperature to 10 K. (b) The similar magnetic relaxation in 100 Oe at 15 K with an intermittent increase in temperature to 20 K.}	
	\label{fig:Fig8}
\end{figure}

It has been discussed earlier that the collective dynamics can be qualitatively explained with the two theoretical models for SG, viz. the droplet model and the hierarchical model. To identify the model which accounts the dynamics in this compound we have collected the $M(t)$ with cycling the temperature. To be specific, in the droplet model scenario the droplets of correlated spins grow with time at particular temperature. With the change in temperature by $\Delta T$, the correlation of spins having spatial order larger than $\ell (\Delta T)$ breaks and new relaxation starts at temperature $T \pm \Delta T$. However, this process is symmetric with the direction of $\Delta T$. On the contrary, in hierarchical model as the metastable states either split or merge with the decrease or increase in temperature respectively, thus the direction of $\Delta T$ should have asymmetric effect. Based on these arguments, the measured $M(t)$ with positive and negative temperature cycle can be helpful to identify the appropriate model for the system.\cite{lefloch,sun} In Fig. 8a, we present the relaxation data following a protocol where the sample has been cooled in zero field from the room temperature to 15 K. After stabilization of temperature, a magnetic field of 100 Oe is applied and $M(t)$ is measured for time $t_1$. Then, without changing the field the sample is cooled to 10 K and $M(t)$ is recorded for time $t_2$, and then the sample is heated back to 15 K and $M(t)$ is measured for the time $t_3$. We used $t_1 = t_2 = t_3$ = 4000 s. It is evident in figure that $M(t)$ increases continuously at 15 K during $t_1$, but the relaxation at 10 K during $t_2$ is almost halted. However, system resumes its relaxation again at 15 K during $t_3$. Moreover, it is observed in figure that the magnetic relaxation at 15 K during $t_3$ is nearly a continuation of that during $t_1$. Similarly, we have measured $M(t)$ for the positive $\Delta T$ where the intermittent temperature is increased to 20 K and data are shown in Fig. 8b. It is observed that magnetization relaxes both at 15 and 20 K but when the system is cooled back to 15 K, the relaxation is almost halted and $M(t)$ during $t_3$ is not a continuation of that during $t_1$. 

This asymmetrical behavior of magnetic relaxation with the positive and negative temperature cycle is quite intriguing. The droplet model being symmetric with the direction of temperature change, we attempt to explain this behavior with the hierarchical model. Following this, at lower temperature the metastable states are divided, therefore the relaxation occurs among the newly born sub states because the barrier between the original states increases with lowering the temperature. When the system is heated back to the original temperature the initial states are recovered with the unchanged relative occupation. As a result, the magnetization during $t_3$ is almost continuation of $t_1$ at same temperature. However, this is not the case for the positive $\Delta T$ as during the relaxation at higher temperature, the relative occupation of states are changed because they are merged into the new state. Consequently, when the system is cooled back to original temperature, the $M(t)$ does not match with the earlier value. The asymmetric behavior in magnetic relaxation with the temperature change in Fig. 8 clearly imply that the dynamics in PSMGO is well explained by the hierarchical model. Nonetheless, this observation is quite intriguing for the PS manganites where the models were originally proposed for the SG dynamics.
 
\maketitle\section{Conclusion}
In conclusion, we have studied the low-$T$ dynamics in Pr$_{0.5}$Sr$_{0.5}$Mn$_{0.925}$Ga$_{0.075}$O$_{3}$, a manganite around half-doping. This compound exhibits the PS phenomenon where the FM and AFM clusters coexist and the size of the former is substantially reduced due to the substitution of nonmagnetic Ga at Mn-site. A new magnetic anomaly appears at low temperature which is marked by a peak in ZFC magnetization. From the detailed measurements of dc magnetization as well as linear and nonlinear ac susceptibility we infer that this peak at low temperature arises due to the thermal blocking of nano-meter size FM clusters exhibiting SPM behavior. It is, however, intriguing that all the experimental features related to collective dynamics and glass-like freezing e.g. slow magnetic relaxation, aging effect, memory effect in both FC and ZFC magnetization are evident in this compound. Based on these results we conclude that the presence of collective behavior which arises from the intercluster interaction in PSMGO. Moreover, the asymmetric behavior of magnetic relaxation with the positive and negative temperature cycle imply that the dynamics in this compound can be explained by invoking the hierarchical model rather than the droplet model which are commonly used to describe the collective dynamics in glassy systems. We believe that this study would stimulate the theoretical and experimental investigations to understand the dynamics in PS system at low temperature, in particular with intercluster interaction.

\maketitle\section{Acknowledgment}
We thank Kranti Kumar for the help in measurements. DST, Government of India is acknowledged for funding VSM. AKP also acknowledges CSIR, India for financial assistance.

\end{document}